# Design and Execution Challenges for Cybersecurity Serious Games: An Overview

Gokul Jayakrishnan, Vijayanand Banahatti, Sachin Lodha
*TCS Research, Tata Consultancy Services, Pune, India*

**Abstract**

Serious games are increasingly being used in cybersecurity education to engage and educate users. Several studies with cybersecurity serious games have shown that they are successful in educating users and the users also find them both fun and engaging. Meanwhile, several studies have also reported issues in identifying real life effects of the game and even the long-term effects that they have. Based on our experience with enterprise cybersecurity games and games from recent literature, we discuss a few key challenges that must be considered while designing and evaluating serious games for cybersecurity awareness.

## 1. Introduction

Serious games are computer games created for the primary purpose of education rather than entertainment [1]. They are designed for training users on a particular skillset [2]. Zyda [3] defines serious games as games with specific rules that use entertainment to further training, education, policies, and strategic communication objectives. Serious games have been used in training in various areas such as healthcare and medical education [4, 5], language learning [10], training cultural heritage [6, 7], academia [8, 9], and in various other fields.

In this paper, we surveyed a set of 15 serious games used for cybersecurity education in academia and enterprise, from 2017 to 2023 (Table 1), with inclusion of Anti-phishing Phil [11] from 2007 as well. The methodology for gathering the games was from Google Scholar Database using a combination of keywords such as "serious games", "games for cybersecurity", "phishing awareness", "enterprise", and "password security awareness". This was similar to the method used by a previous survey paper [39], which compared several cybersecurity serious games. We focused more on computer-based serious games, with corresponding studies conducted and papers published in academic conferences primarily related to usable security and privacy and human-computer interaction (HCI). However, while [39] was a general survey paper, we wanted to suggest possible challenges in this area towards the future direction. Considering the vast array of cybersecurity serious games being used in literature, and with the likelihood that serious games will be used more soon, we explore our research question: What are the key challenges to look out for while designing cybersecurity serious games? In the following sections, we outline certain major challenges that were observed from our literature survey, and we also suggest some interventions to some of the challenges.

## 2. Literature

There is a major body of literature that deals with phishing and password security-related games. We observed some more recent studies involving serious games and identified the areas that could be more focused to help get a better learning outcome.

i. **Anti-phishing Phil (2007):** Among the serious games primarily used for cybersecurity education, Anti-phishing Phil [11] is one of the most widely cited and well-known studies in literature. This is an anti-phishing awareness game developed by Carnegie Mellon University (2007) that teaches users good habits and help them avoid phishing attacks. We considered this game in our literature survey because it can be considered as one of the games that laid certain foundations for cybersecurity serious games research. The game had a control condition and game condition having pre-test and post-test in which the latter showed better scores for the game suggesting the effectiveness of game over traditional methods of training.

ii. **PASDJO (2017)**: This is a game where players analyze strengths of various passwords under time pressure [18]. The participant data for four months show that users often underestimate passphrases. However, the paper also reports that strength estimation is learnable and the more the participants play the game, the more they learn. The authors suggest that the feedback mechanism in the game could have helped people learn more.

iii. **Bird's Life (2018):** This 2D game [17] was intended for college-level students to learn about anti-phishing. The game was played by 100 students from both CS/IT and non-CS/IT courses. Its assessment involved a pre-test and post-test, in which the post-test showed higher performance, showing an impact of the game. The non-CS/IT group of participants showed a 67% improvement which is very high. The participant feedback from 138 participants shows the game was very favorable and educational.

Table 1: Brief overview of 15 games reviewed.

| Number | Game | Area Focused | Analysis Method | Audience | Participant Count |
|---|---|---|---|---|---|
| 1 | Anti-phishing Phil [11] | Phishing | Pre-test-post-test, Control condition | Students | 14 (Initial study) |
| 2 | PASDJO [18] | Passwords | Scoring Algorithm | Students | 115 |
| 3 | Bird's Life [17] | Phishing | Pre-test-post-test, Online survey | College Students | Over 100 |
| 4 | Phishy [12] | Phishing | Pre-test-post-test | Employees | 8,071 |
| 5 | Pomega [15] | Phishing, Passwords, Mobile Security, Network Security, Physical Security | Pre-test-post-test | Students | Three user groups |
| 6 | Interland [16] | Phishing, Anti-bullying, online safety, password strength | Game Analysis | K-12 | Data not available |
| 7 | GenCyber [36] | Social engineering, Information Security, Online safety and Cybersecurity principles | Game and Post-task questionnaire | High School Students | 181 |
| 8 | What.Hack [21] | Phishing emails | Pre-test-post-test, Control condition | Students | 13 |
| 9 | RPG game [25] | Password Security | Qualitative analysis | General users | 17 |
| 10 | Passworld [13] | Passwords | Pre-test-post-test | Employees | 4,906 |
| 11 | Riskio [20] | Attack vs Defense in security | Game and Post-task questionnaire | Employees, University Students | 54 |
| 12 | PickMail [19] | Phishing emails | Pre-test-post-test | Employees | 478 |
| 13 | SherLOCKED [23] | Foundational Security Concepts | Game- Post-test survey | Students | 112 |
| 14 | Housie (multi-player) [24] | Network Security, Application Security, Operational Security, End user education and Information Security | Game Analysis | Employees | 56,358 |
| 15 | Secure Workspace [22] | Work-from-home Security | Pre-test-post-test | Employees | 36,390 |

iv. **Phishy (2018):** Phishy is an anti-phishing awareness game that was launched for an enterprise audience [12]. The authors used a pre-test-post-test methodology to analyze the learning outcome of the game with 8071 participants, in which the post-test showed a higher performance as compared to the pre-test. The study also reports that the increase in knowledge was found to be more in participants who had lower pre-test scores.

v. **Pomega (2018):** A 2D game [15] for promoting cybersecurity knowledge, covering five topics phishing, password, social network, mobile security, and physical security. The game was evaluated by three user groups, and the results revealed that all users gained better post-test scores than the pre-test scores and found the game engaging.

vi. **Interland (Google) (2018):** As a serious game designed for students to be safe in the cyberworld [16], the topics focused were anti-bullying, content safety while posting online on phishing and password strength. Students were awarded a certificate after completion of the game.

vii. **GenCyber Camp (2018):** Jointly funded by the National Security Agency and the National Science Foundation, this innovative game-based learning camp with four games [36] was developed for educating high school (K-12) students on concepts of social engineering, information security, safe online behavior, and several cybersecurity principles. This study found improvement in cybersecurity behavior of students post the game and found that the male students enjoyed the game more than the female students.

viii. **What.Hack (2019):** To teach phishing concepts and to simulate actual phishing attacks in a role-player setting, the online serious game What.Hack was developed [21], which is both engaging and educational. The participants interact with several simulated emails and try to judge their legitimacy, with appropriate feedback helping them learn in the process. The game was compared to a standard form of training and another anti-phishing serious game, where What.hack was found to be more engaging and effective in improving participants' performance.

ix. **RPG Quiz on Passwords (2019):** To educate on password security, Scholefield et. Al [25] developed a role-playing game (RPG) on android platform. The ensuing study with 17 participants found that the users enjoyed learning while using the application and the qualitative study suggests that the participants benefited from the gamification.

x. **Passworld (2020):** This is an enterprise-oriented serious game to train employees in the concepts of password security [13]. Passworld is a platformer game [14] where the player controls a game character to navigate through various 2-dimensional platforms and gaining password security-related information in the process. Passworld also had a pre-test and post-test, in which the latter showed better performance improvement among the enterprise audience. The study details certain limitations such as lack of a control condition and inability to test the effectiveness of the game in a real-world setting, including the test of long-term password memorability.

xi. **Riskio (2020):** Riskio [20] is a tabletop card-based serious game to increase cybersecurity awareness among people with no technical background, working in organizations. Here, the players gain knowledge of cybersecurity attacks and defenses by actively taking the role of attacker and defender in a fictitious organization. The evaluation of the game shows that employees have higher confidence than students that Riskio can increase their awareness on cybersecurity concepts.

xii. **PickMail (2022):** This is an email-based anti-phishing awareness game [19] teaching users how to identify various hints within emails to judge their legitimacy. The study was launched in an enterprise setting with 478 participants and showed higher post-test scores (92.6%) as compared to pre-test scores (76.8%).

xiii. **SherLOCKED (2021):** This is a 2D, top-down puzzle adventure game used to consolidate students' knowledge of foundational security concepts [23]. It was implemented within undergraduate course and tested with 112 participants (students) and the study reports the game to be effective, attractive, and allowing further engagement with the contents and over 87% participants felt that the game helped improve the understanding of lecture content.

xiv. **Housie (2022)**: A BINGO-themed multiplayer serious game for equipping enterprise employees with cybersecurity knowledge [24]. The game was played by over 56000 participants in the enterprise sector and the study found that the multiplayer format was widely accepted, and the game was found to be engaging and entertaining.

xv. **Secure Workspace (2023)**: This game [22] focused on gauging security awareness of enterprise employees and simultaneously providing them with conceptual training on the several dos and don'ts of working from home, including physical device security and unauthorized disclosure of information. The game was played by over 36000 participants and the paper reports that the enterprise employees performed well in identifying various security violations and physical security is one area that needs further training. The paper also suggests various interventions that can help reduce security violations during work-from-home.

We consolidated a list of 15 games here, however, we could not find any evidence for a long-term effect of retaining cybersecurity knowledge or participants effectively carrying forward the knowledge gained into real life situations. We surveyed these (and a few other) games; based on our analysis of the findings and limitations, and also based on our experience developing enterprise serious games, we came up with a list of potential challenges such as difficulties in long term knowledge retention, identifying effects of learning in real life, nature of demographics of audience etc. While performing the literature survey, we found that some of these challenges also apply to serious games from academia as well. In the following section, we put forward the potential challenges we identified to educate the future creators of cybersecurity serious games, to incorporate as many solutions as possible for the said challenges.

## 3. Discussion on Challenges of Cybersecurity Serious Games

From the studies of existing serious games on cybersecurity, we found certain common limitations and challenges that could affect real-time testing of the effects of the games, which are given below:

**1) Knowledge Retention:** While several papers have shown a relatively higher participant performance in the post-tests, their long-term knowledge retention is not studied. A recent role-playing quiz-based serious game performed a retention study after five months [25], on participants' knowledge on phishing links and found that even though they performed well, their overall performance decreased, suggesting issues with knowledge retention. Use of long-term retention studies after a period of more than two months using tests and/or questionnaires could help the researchers get a better idea of knowledge retention among the participants.

**2) Participation:** A study requires wholehearted participation from the study subjects. In case of a long-term security awareness study featuring several steps such as a pre-test, a training condition such as a serious game, and a post-test, the participant must ensure they complete each of the above without fail. Encouraging participants to complete all stages of the experiment, especially the long-term knowledge retention test, should be mandated to collect valuable data for analysis, that could in turn provide with insights on the enterprise user behavior and trends. However, it should also be noted that forcing participants to comply would result in them considering this as extra work [35], which might also not provide a positive outcome.

**3) Social Desirability Bias:** This is another major reason for the difficulty in identifying honest responses from the participants. During our literature survey, we came across a serious game, in virtual reality, used as an interpretive tool for tourism of the Great Barrier Reef, Australia [26]. Despite the game having a positive response and having shown that the tourists developed positive emotions towards the threatened ecosystem, the authors doubt that social desirability bias might be in play here. During surveys and tests, there is a small chance that the participants might conceal their actual opinion and respond in such to make themselves align with the best possible answers [33]. This will do more harm to the study since this might hide the true responses of the participants and the analysis may get too much positive responses. This could shift the correctness percentages both in the pre-test and post-test.

**4) Real-life Effects**: Most of the cybersecurity serious games from literature exist as standalone games that are separate from the normal daily activities of the participants, especially the students or employees. Therefore, attempting these games might provide a sense of alertness, which could also give rise to social desirability bias and extra caution. The same participants might not do so in a real-life situation. Identifying the steps to measure the real-world learning outcome from the serious games is also a major challenge to tackle. Based on our experience with enterprise serious games [12, 13, 19, 22, 24] their real-life effects are difficult to study in the enterprise setting. One suggestion for future serious games on cybersecurity is to integrate the teaching and the game within the confines of the participants' daily activities so as not to alarm them and this could also help obtain more genuine responses.

**5) Nature of Audience**: The studies involving enterprise games [12,13,19, 22] have shown their demographics as a limitation, which mainly consists of enterprise employees, who might have had certain security training in the past and are more likely to answer security-related questions much better than those who have had no such training. Studies using serious games could focus on a wide range of demographics, with focus to participants having both CS/IT and non-CS/IT knowledge. If possible, extending the study to a general audience could help gather more important data regarding the cybersecurity serious games, and could even result in much higher performance increase. An interesting demographics could also be what is called as 'emergent users' [37], people who are just about to access advanced mobile devices and services, especially from developing countries, and therefore are more likely to encounter cybersecurity threats and attacks.

**6) Life-like Simulations**: In real life, checking an email and judging them usually happens within a short span of time, with one study showing an average of 15-20 seconds per email [34]. The games that simulate phishing emails and even password entry should ensure that appropriate time-based motivators should be provided to the participants while responding to emails/passwords. Therefore, simulating real life scenarios would be an ideal method to promote users to learn quick thinking and performing appropriate actions.

**7) Time:** Spending too much time on cybersecurity serious games might also be uneventful if the users find them to be boring after a certain time. For enterprise, if the employees find the games to be an extra work or a hindrance could result in negative effects [35]. Developers of cybersecurity serious games must ensure that the games do not become too lengthy to keep the audience engaged and to reduce the effects of fatigue. They should also not be too short as the participants would need sufficient time to understand and learn the game mechanics prior to interacting with the game for learning.

**8) Control Condition and Analysis:** With hundreds of participants taking part in cybersecurity serious games being conducted at academic or corporate locations, analysis of their data is very important in assessing the effectiveness of the game. Performance in pre-test and post-test, along with control conditions are the general approaches done by studies so far. Could a real-time simulation provide more accurate data for analysis? Does the demographics play an important role in cybersecurity education through serious games? One study says no [11] and other study says it does [12]. Is there a consensus here? Future serious games should look at more holistic methods to identify the effectiveness of the game, considering all the factors that are dependent on the learning outcome.

**9) Repurposability:** 'Repurposability' [38] or the ability to get repurposed for multiple learning contexts can be a very useful design decision for upcoming cybersecurity serious games. The development time can be lowered to great extends and the learning content can be improved and altered within a short period, which would enable cybersecurity serious games to be tested quickly and efficiently.

## 4. Conclusion

We analyzed several recent serious games for cybersecurity training and awareness and identified several common challenges that, if rectified, could likely result in a better learning outcome and assessment of the serious games. Effective ways for ensuring knowledge retention, avoiding biases, increasing participation, and efficient data analysis are a few of them. While we consider this survey to be limited, focusing on just 15 games, we plan to explore further and include a larger set of games for our detailed analysis. We hope to gather more insights about the current state of serious games and their challenges and hope to provide possible solutions for at least a few of them based on existing studies.